\documentstyle[12pt] {article}
\newcommand{\beq}{\begin{equation}}
\newcommand{\eeq}{\end{equation}}
\newcommand{\beqa}{\begin{eqnarray}}
\newcommand{\eeqa}{\end{eqnarray}}
\newcommand{\ba}{\begin{array}}
\newcommand{\ea}{\end{array}}

\begin{document}

\begin{center}
{\large \bf Quantum Chaos in Vibrational Nuclei: \\
the Green Function Approach}
\footnote{This work has been partially supported by the Ministero
dell'Universit\`a e della Ricerca Scientifica e Tecnologica (MURST).}
\end{center}

\vspace{0.5 cm}

\begin{center}
{\bf V.R. Manfredi$^{(a)}$, M. Rosa--Clot$^{(b)(c)}$\\
L. Salasnich$^{(d)}$ and S. Taddei$^{(b)(c)}$}
\vskip 0.5 truecm
$^{(a)}$ Dipartimento di Fisica ``G. Galilei" dell'Universit\`a
di Padova, \\
INFN, Sezione di Padova, \\
Via Marzolo 8, I--35131 Padova, Italy
\\
Interdisciplinary Laboratory, SISSA,\\
Strada Costiera 11, I--34014 Trieste, Italy

\vskip 0.5 truecm

$^{(b)}$ Dipartimento di Fisica dell'Universit\`a di Firenze, \\
Largo E. Fermi 2, I--50125 Firenze, Italy

\vskip 0.5 truecm

$^{(c)}$ INFN, Sezione di Firenze, \\
Largo E. Fermi 2, I--50125 Firenze, Italy

\vskip 0.5 truecm

$^{(d)}$ Departamento de Fisica Atomica, Molecular y Nuclear \\
Facultad de Ciencias Fisicas, Universidad Complutense de Madrid, \\
Ciudad Universitaria, E--28040 Madrid, Spain
\end{center}

\newpage

\begin{center}
{\bf Abstract}
\end{center}
\vskip 0.5 truecm
\par
We show that the collective vibrational model of atomic nuclei
displays quantum chaos. To avoid the problems related to the tunneling
effects, a Green function deterministic numerical method has been used to
evaluate the energy levels.

\newpage

{\bf 1. Introduction}
\vskip 0.5 truecm
\par
In the last years many studies have been devoted to the properties
of quantal systems which are chaotic in the classical limit [1].
\par
In quantum mechanics, given the impossibility of defining the
trajectories, classical concepts and methods cannot be applied.
Nevertheless, many
efforts have been made to establish the features of quantum systems
which ref\/lect the qualitative difference in the behaviour of their
classical counterparts. Many schematic models have shown that
this difference reveals itself in the properties of fluctuations in
eigenvalue sequences. The spectral statistics for the systems with
underlying classical chaotic behaviour agree with the predictions of
the random matrix theory.
By contrast, quantum analogs of classically integrable systems
display the characteristics of Poisson distribution [2,3].
\par
In atomic nuclei, the experimental data of nuclear spectroscopy
suggest regular states near the yrast line and
chaotic states near the neutron emission threshold, but
the coexistence of regular, chaotic and collective
states is not yet well understood [4,5].
\par
In a previous paper [6] we studied the transition form order to
chaos in the nuclear roto--vibrational model [7,8],
while in this work we analyze in greater detail the numerical technique
to obtain the energy levels and the effects of
the order--chaos transition in the spectral statistics
for the collective vibrational motion.

\newpage

\par
{\bf 2. The model}
\vskip 0.5 truecm
\par
The two--dimensional model which describes the collective vibrational
motion of atomic nuclei was introduced by Bohr and Mottelson [7]
and developed by Eisenberg and Greiner [8].
The Hamiltonian is given by
\beq
H={1\over 2}B({\dot x_1}^2+2{\dot x_2}^2)+V(x_1,x_2),
\eeq
where
\beq
V(x_1,x_2)={1\over 2}C_2(x_1^2+2x_2^2)+
\sqrt{2\over 35}C_3x_1(6x_2^2-x_1^2)+{1\over
5}C_4(x_1^2+2x_2^2)^2+V_0.
\eeq
The variables $x_1$ and $x_2$ are connected to the deformation
$\beta$ and asymmetry $\gamma$ by the standard relations [9]
\beq
x_1=\beta \cos{\gamma}, \;\;\; x_2={\beta\over \sqrt{2}}\sin{\gamma}.
\eeq
As discussed in [8], the presence of bound states in atomic nuclei leads
to a value of $C_4>0$, whereas for $C_3$ a positive value corresponds to
a prolate shape, a negative value to an oblate shape. Similarly $C_2$ may
also be either positive or negative.
\par
The shape of the nuclear potential $V(x_1,x_2)$ is a function of $C_2$
and $\chi =C_3^2/(C_2 C_4)$. For $C_2>0$, and $0<\chi<56/9$
the nucleus is spherical; for $56/9<\chi<7$ the nucleus is spherical in
the ground state (g.s. spherical) and deformed in the excited states
(e.s. deformed); for $\chi >7$ it is g.s. deformed and e.s. spherical;
for $C_2<0$ it is g.s. deformed and $\gamma$--unstable in the excited
states.
\par
The chaotic behaviour of this model has been studied analytically
by the authors of [9] using the criterion of negative curvature of
the  potential energy. However, this criterion has been demonstrated
to be inaccurate in [10].
The authors of [9] have also obtained the energy
levels by quantizing the Hamiltonian (1) for g.s. spherical nuclei.
For g.s. deformed nuclei, like rare--earth, however, it is not simple
to obtain the correct energy levels because the potential energy
is an asymmetric triple well.
\par
Table 1 shows the parameters of the nuclear potential (2), for the
rare--earth $^{160}Gd$ and $^{166}Er$ according to [8].
The potential energy for these nuclei is plotted in Figure 1.

\vskip 0.5 truecm
\par
{\bf 3. The GFND method}
\vskip 0.5 truecm
\par
To obtain the energy levels, we use the Green function numerical
diagonalization method (GFND) [11--14], whose
starting point is the integral Schr\"odinger equation
\begin{equation}
\int\int
K(x_1,x_2,y_1,y_2;\varepsilon)\psi_n(y_1,y_2)dy_1dy_2=
e^{-\varepsilon E_n}\psi_n(x_1,x_2).
\end{equation}
$K(x_1,x_2,y_1,y_2;\varepsilon)$ is the Euclidean short
time propagator and $\psi_n(x_1,x_2)$ are the Hamiltonian
eigenfunctions. This equation can be approximated using a numerical
integration rule which gives
\begin{equation}
\sum_{h=1}^N\sum_{k=1}^Nw_{1,h}w_{2,k}\;K^{\varepsilon}_{ijhk}\;
\psi^n_{hk}\simeq e^{-\varepsilon E_n}\; \psi^n_{ij}.
\end{equation}
$\psi^n_{ij}\equiv\psi_n(x_{1,i},x_{2,j})$,
$K^{\varepsilon}_{ijhk}\equiv
K(x_{1,i},x_{2,j},x_{1,h},x_{2,k};\varepsilon)$, $w_{1,i}$,
$w_{2,i}$ are the weights associated with the integration rule
(a good choice is the trapezoidal one). The intervals of integration
in Eq. (4) go from $-\infty$ to $+\infty$, but in Eq. (5) we take finite
matrices and consequently finite intervals $L_1$ and $L_2$.
This corresponds to confining the system in a ``box'' of sides $L_1$ and
$L_2$. However, if the intervals are large enough, this gives
negligible corrections to the energies and to the wave functions of
the bound states. Therefore, by diagonalizing the matrix
$w_{1,h}w_{2,k}\;K^{\varepsilon}_{ijhk}$, we obtain the
energies and the wave functions directly.
\par
The Euclidean short time propagator can be written in the form (we
use units \hbox{$B=\hbar=1$}):
\begin{equation}
K(x_1,x_2,y_1,y_2;\varepsilon)={1\over 2\pi\varepsilon}\exp\left\{ -{1\over
2\varepsilon}[(x_1-y_1)^2+2(x_2-y_2)^2]-f(x_1,x_2,y_1,y_2;\varepsilon)
\right\},
\end{equation}
where the first term in the exponential corresponds to the kinetic
part of the Hamiltonian and the function
$f(x_1,x_2,y_1,y_2;\varepsilon)$ is the potential term.
Its explicit expression depends on the prescription chosen; for
example the last point rule
\hbox{$f(x_1,x_2,y_1,y_2;\varepsilon)=\varepsilon V(x_1,x_2)$} gives a
propagator which is correct up to $O(\varepsilon)$ only, while the symmetric
expression \hbox{$\displaystyle{{\varepsilon\over 2}}\;
[V(x_1,x_2)+V(y_1,y_2)]$} is correct up to $O(\varepsilon^2)$.
Moreover, a systematic expansion of the short time propagator in
$\varepsilon$, $\Delta_1=x_1-y_1$ and $\Delta_2=x_2-y_2$ is also
possible [13]. This expansion gives rise to an asymptotic series,
which, if used within the GFND method with
non--singular potentials, allows one to obtain high numerical accuracy.
There follows a brief sketch of the expansion method.
\par
Since the potential term of $K(x_1,x_2,y_1,y_2;\varepsilon)$ is
weighed by the Gaussian function: $\exp\left\{ -\displaystyle{{1\over
2\varepsilon}}[(x_1-y_1)^2+2(x_2-y_2)^2]\right\}$, the difference
$\Delta_i$ can be considered of the order $\sqrt{\varepsilon}$. As a
consequence, we can expand in $\Delta_1$ and $\Delta_2$ as well as in
$\varepsilon$. Therefore, following Ref. [13], let us write the short
time propagator in the form
$$
K(x_1,x_2,y_1,y_2;\varepsilon)=
{1\over 2\pi\varepsilon}\exp\{ -{1\over 2\varepsilon}[(x_1-y_1)^2+2(x_2-y_2)^2]
$$
\begin{equation}
-\sum_{\nu=0}^N\sum_{\mu=0}^{2(N-\nu)}
\sum_{i_1=1}^2\!\ldots\!\sum_{i_\mu=1}^2
g_{\mu\nu}^{i_1,\ldots,i_\mu}(y_1,y_2)\;
(x_{i_1}-y_{i_1})\ldots(x_{i_\mu}-y_{i_\mu})\varepsilon^\nu\}.
\end{equation}
By expanding up to $O(\varepsilon^4)$ and requiring that
$K(x_1,x_2,y_1,y_2;\varepsilon)$ satisfies the Schr\"odinger equation, we
obtain the coefficients $g_{\mu\nu}^{i_1,\ldots,i_\mu}$
(see Appendix 1).
\par
The GFND gives accurate results and is more precise than
techniques based on the direct diagonalization of the Hamiltonian,
especially in the case of tunneling problems
(for a deeper discussion of this point see [11--13]).
Therefore, it is an appropriate method of studying the potential
given in Eq.\ (2), where we have small tunneling effects between wells.

\vskip 0.5 truecm
\par
{\bf 4. Numerical results}
\vskip 0.5 truecm
\par
As is well known, the classical global instability of a system
is appropriately studied using Poincar\'e sections.
Regular regions are characterised by sets of invariant intersection
points; chaotic regions by points which are
distributed irregularly [15].
\par
{}From the quantal point of view, the energy spectrum obtained using the
GFND, has quasi--degenerate levels, due to the presence of three
similar potential wells, two of which are identical. Therefore, it
would be necessary to separate the energy levels into three classes,
corresponding to the states localized in each well. For our purpose,
however, we need only take one of the two double degenerate
levels, which are easily identified. Thus, we obtain approximately one
hundred energy levels, which are enough for statistical significance.
The energy spectrum thus obtained has been mapped into one
with a quasi-uniform level density, by performing the unfolding
procedure  described in detail in reference [16].
Then, the distribution $P(s)$ of spacings between adjacent levels and the
spectral rigidity $\Delta_3(L)$ [17,18] has been
calculated. These spectral statistics are
compared to Poisson statistics $P(s)=e^{-s}$ and
$\Delta_3(L)=L/15$ of integrable systems, and to
Gaussian Orthogonal Ensemble (GOE) statistics
$P(s)={\pi\over 2}s e^{-s^2\pi/4}$ and
$\Delta_3(L)=\pi^{-2}\ln{(L)}-0.0007$ of chaotic systems.
\par
Fig. 2 shows that in $^{160}$Gd, which has a high saddle energy ($5.56$ MeV),
there is coexistence of regular and chaotic motion
for energies below the saddle energy.
In this region, the system has only a few energy levels and so we
cannot perform a good statistical study of quantum levels.
\par
Fig. 3 and Fig. 4 show that in $^{160}$Gd
above the saddle energy, at about $8$ MeV,
there is chaos. For higher energies, at about $18$ MeV,
there is a quasi--regular behaviour.
The non--universal behaviour of $\Delta_3(L)$
for high values of $L$, not predicted by GOE, has been explained by
Berry [19] using semiclassical quantization.
\par
Fig. 5 and Fig. 6 show that in $^{166}$Er, which has a low saddle energy
($1.89$ MeV), for energies above the saddle energy, at about $4$ MeV, there is
prevalently chaotic behaviour; for higher energies, at about $15$ MeV,
there is a predominance of regular classical
trajectories.

\vskip 0.5 truecm
\par
{\bf Conclusions}
\vskip 0.5 truecm
\par
In conclusion, the collective vibrational behaviour of the above
nuclei displays a quasi--chaos $\to$ quasi--order
transition as a function of the energy. The mixed behaviour
at very low energies cannot be easily shown by
spectral statistics because only a few energy levels are present.
\par
For the sake of completeness we observe that other phenomenological
collective models have been recently proposed, see for example [20] and
references therein.

\vskip 0.5 truecm
\begin{center}
{\bf Acknowledgments}
\end{center}
\vskip 0.5 truecm
\par
L.S. has been supported by a Fellowship from the University of Padova
and is grateful to the "Ing. Aldo Gini" Foundation of Padova
for partial support. V.R.M is very grateful to Prof. S. Fantoni,
Director of the Interdisciplinary Laboratory, SISSA (Trieste), for his
kind hospitality in his laboratory, where the final version of this
paper was written.

\newpage

\par
{\bf Appendix 1}
\vskip 0.5 truecm
\par
The coefficients $g_{\mu\nu}^{i_1,\ldots,i_\mu}$ of Eq.\ (7) are given
by (the $y_1$ and $y_2$ dependence in $g$ and $V$ is understood)

\begin{eqnarray}
g_{01}&=&V^{(0,0)}\nonumber\\
g_{11}^{i_1}&=&{1\over 2}\;{\partial V\over\partial
x_{i_1}}\nonumber\\
g_{21}^{i_1i_2}&=&{1\over 6}\;{\partial^2
V\over\partial x_{i_1}\partial x_{i_2}}\nonumber\\
g_{31}^{i_1i_2i_3}&=&{1\over 24}\;{\partial^3 V\over\partial
x_{i_1}\partial x_{i_2}\partial x_{i_3}}\nonumber\\
g_{41}^{i_1i_2i_3i_4}&=&{1\over 120}\;{\partial^4 V\over\partial
x_{i_1}\partial x_{i_2}\partial x_{i_3}\partial x_{i_4}}\nonumber\\
g_{51}^{i_1i_2i_3i_4i_5}&=&{1\over 720}\;{\partial^5 V\over\partial
x_{i_1}\partial x_{i_2}\partial x_{i_3}\partial x_{i_4}\partial
x_{i_5}}\nonumber\\
g_{61}^{i_1i_2i_3i_4i_5i_6}&=&{1\over 5040}\;{\partial^6
V\over\partial x_{i_1}\partial x_{i_2}\partial x_{i_3}\partial
x_{i_4}\partial x_{i_5}\partial x_{i_6}}\nonumber\\
g_{02}&=&{1\over
12}\;(V^{(0,2)}+V^{(2,0)})\nonumber\\
g_{12}&=&{1\over 24}\left(
\begin{array}{cc}
                     V^{(1,2)}+V^{(3,0)},&\!\!\!V^{(0,3)}+V^{(2,1)}\\
\end{array} \right)\nonumber\\
g_{22}&=&{1\over 80}\left(
\begin{array}{cc}
                      V^{(2,2)}+V^{(4,0)},&\!\!\!V^{(1,3)}+V^{(3,1)}\\
                      V^{(1,3)}+V^{(3,1)},&\!\!\!V^{(0,4)}+V^{(2,2)}\\
\end{array} \right)\nonumber\\
g_{32}^{111}&=&{1\over 360}\;(V^{(3,2)}+V^{(5,0)})\nonumber\\
g_{32}^{112}=g_{32}^{121}=g_{32}^{211}&=&
{1\over 360}\;(V^{(2,3)}+V^{(4,1)})\nonumber\\
g_{32}^{122}=g_{32}^{221}=g_{32}^{212}&=&
{1\over 360}\;(V^{(1,4)}+V^{(3,2)})\nonumber\\
g_{32}^{222}&=&{1\over 360}\;(V^{(0,5)}+V^{(2,3)})\nonumber\\
g_{42}^{1111}&=&{1\over 2016}\;(V^{(4,2)}+V^{(6,0)})\nonumber\\
g_{42}^{1112}=g_{42}^{1121}=g_{42}^{1211}=g_{42}^{2111}&=&
{1\over 2016}\;(V^{(3,3)}+V^{(5,1)})\nonumber\\
g_{42}^{2211}=g_{42}^{2121}=g_{42}^{2112}\;\phantom{=g_{42}^{1122}}
&\phantom{=}&\nonumber\\
=g_{42}^{1221}=g_{42}^{1212}=g_{42}^{1122}&=&
{1\over 2016}\;(V^{(2,4)}+V^{(4,2)})\nonumber\\
g_{42}^{2221}=g_{42}^{2212}=g_{42}^{2122}=g_{42}^{1222}&=&
{1\over 2016}\;(V^{(1,5)}+V^{(3,3)})\nonumber\\
g_{42}^{2222}&=&{1\over 2016}\;(V^{(0,6)}+V^{(2,4)})\nonumber\\
g_{03}&=&{1\over 24}\;(-V^{(0,1)2}+V^{(0,4)}/10\nonumber\\
&\phantom{=}&\phantom{{1\over 24}(}
-V^{(1,0)2}+V^{(2,2)}/5\nonumber\\
&\phantom{=}&\phantom{{1\over 24}(}
+V^{(4,0)}/10)\nonumber\\
g_{13}^{1}&=&{1\over 24}\;(-V^{(0,1)}V^{(1,1)}+V^{(1,4)}/20\nonumber\\
&\phantom{=}&\phantom{{1\over 24}(}
+V^{(5,0)}/20+V^{(3,2)}/10\nonumber\\
&\phantom{=}&\phantom{{1\over 24}(}
-V^{(1,0)}V^{(2,0)})\nonumber\\
g_{13}^{2}&=&{1\over 24}\;(-V^{(0,1)}V^{(0,2)}+V^{(0,5)}/20\nonumber\\
&\phantom{=}&\phantom{{1\over 24}(}
+V^{(4,1)}/20+V^{(2,3)}/10\nonumber\\
&\phantom{=}&\phantom{{1\over 24}(}
-V^{(1,0)}V^{(1,1)})\nonumber\\
g_{23}^{11}&=&{1\over 10}\;(-V^{(1,1)2}/9-V^{(2,0)2}/9\nonumber\\
&\phantom{=}&\phantom{{1\over 10}(}
-V^{(0,1)}V^{(2,1)}/8\nonumber\\
&\phantom{=}&\phantom{{1\over 10}(}
-V^{(1,0)}V^{(3,0)}/8+V^{(2,4)}/168\nonumber\\
&\phantom{=}&\phantom{{1\over 10}(}
+V^{(4,2)}/84+V^{(6,0)}/168)\nonumber\\
g_{23}^{12}=g_{23}^{21}&=&{1\over
10}\;(-V^{(0,2)}V^{(1,1)}/9\nonumber\\
&\phantom{=}&\phantom{{1\over 10}(}
+V^{(1,5)}/168-V^{(1,1)}V^{(2,0)}/9\nonumber\\
&\phantom{=}&\phantom{{1\over 10}(}
-V^{(1,0)}V^{(2,1)}/8+V^{(3,3)}/84\nonumber\\
&\phantom{=}&\phantom{{1\over 10}(}
-V^{(0,1)}V^{(1,2)}/8+V^{(5,1)}/168)\nonumber\\
g_{23}^{22}&=&{1\over 10}\;(-V^{(1,1)2}/9\nonumber\\
&\phantom{=}&\phantom{{1\over 10}(}
-V^{(0,2)2}/9-V^{(1,0)}V^{(1,2)}/8\nonumber\\
&\phantom{=}&\phantom{{1\over 10}(}
-V^{(0,1)}V^{(0,3)}/8+V^{(4,2)}/168\nonumber\\
&\phantom{=}&\phantom{{1\over 10}(}
+V^{(2,4)}/84+V^{(0,6)}/168)\nonumber\\
g_{04}&=&{1\over 10}\;(-V^{(0,2)2}/36+V^{(6,0)}/672\nonumber\\
&\phantom{=}&\phantom{{1\over 10}(}
-V^{(0,1)}V^{(0,3)}/12+V^{(0,6)}/672\nonumber\\
&\phantom{=}&\phantom{{1\over 10}(}
-V^{(1,1)2}/18-V^{(1,0)}V^{(1,2)}/12\nonumber\\
&\phantom{=}&\phantom{{1\over 10}(}
-V^{(2,0)2}/36-V^{(0,1)}V^{(2,1)}/12\nonumber\\
&\phantom{=}&\phantom{{1\over 10}(}
+V^{(2,4)}/224-V^{(1,0)}V^{(3,0)}/12\nonumber\\
&\phantom{=}&\phantom{{1\over 10}(}
+V^{(4,2)}/224),
\end{eqnarray}
where $\displaystyle{V^{(n,m)}={\partial^2 V\over\partial x^n\partial y^m}}$.

\newpage

{\bf REFERENCES}
\vskip 0.5 truecm

[1] M. C. Gutzwiller, {\it Chaos in Classical and Quantum Mechanics}
\break(Springer--Verlag, Berlin, 1990)

[2] A. M. Ozorio de Almeida, {\it Hamiltonian Systems: Chaos and
Quantization}, Cambridge University Press (1990)

[3] K. Nakamura, {\it Quantum Chaos}, Cambridge Nonlinear Science
Series  (1993);
{\it From Classical to Quantum Chaos}, SIF Conference Proceedings, vol.
{\bf 41}, Ed. G. F. Dell'Antonio, S.
Fantoni, V. R. Manfredi (Editrice Compositori, Bologna, 1993)

[4] O. Bohigas and H. A. Weidenm\"uller, Ann. Rev. Nucl. Part.
Sci. {\bf 38}, 421 (1988)

[5] M. T. Lopez--Arias, V. R. Manfredi and L. Salasnich,
Riv. Nuovo Cimento {\bf 17}, 1 (1994)

[6] V. R. Manfredi and L. Salasnich, Int. J. Mod. Phys. E {\bf 4}, 625 (1995)

[7] A. Bohr, B. Mottelson, {\it Nuclear Structure}, vol. 2 (Benjamin,
London, 1975)

[8] J. M. Eisenberg, W. Greiner, {\it Nuclear Models}, vol. 1
(North Holland, Amsterdam, 1970)

[9] Yu. Bolotin, V. Yu. Gonchar, E. V. Inopin, V. V. Levenko, V. N.
Tarasov and  N. A. Chekanov, Sov. I. Part. Nucl. {\bf 20}, 372
(1989)

[10] G. Benettin, R. Brambilla and L. Galgani, Physica A {\bf
87},  381 (1977)

[11] M. Rosa--Clot and S. Taddei, Phys. Rev. C {\bf50}, 627 (1994).

[12] M. Rosa--Clot and S. Taddei, Proceedings of the
``V Convegno su Problemi di Fisica Nucleare Teorica'', Cortona 1993.

[13] M. Rosa--Clot and S.Taddei, Phys. Lett. A {\bf 197}, 1 (1995).

[14] S. Taddei, submitted to Phys. Rev. C.

[15] H. Poincar\`e, {\it New Methods of Celestial Mechanics}, vol. 3,
ch.  27 (Transl. NASA Washington DC 1967);
M. Henon, Physica D {\bf 5}, 412 (1982)

[16] V. R. Manfredi, Lett. Nuovo Cimento {\bf 40}, 135 (1984)

[17] F. J. Dyson, M. L. Mehta: {\it J. Math, Phys.} {\bf 4}, 701 (1963)

[18] O. Bohigas, M. J. Giannoni: Ann. Phys. (N.Y.) {\bf 89}, 393 (1975)

[19] M. Berry, Proc. Roy. Soc. Lond. A {\bf 400}, 229 (1985)

[20] A. A. Raduta, V. Baran, D. S. Delion, Nucl. Phys. A {\bf 588}, 431 (1995)

\newpage

{\bf TABLE CAPTIONS}
\vskip 0.5 truecm

Table 1: Numerical values of the parameters $C_2$, $C_3$, $C_4$, $V_0$
defining the nuclear potential $V$ for the two rare--earth nuclei
studied.

\newpage

{\bf FIGURE CAPTIONS}
\vskip 0.5 truecm

Figure 1: The potential energy of the rare--earth $^{160}Gd$ (a) and
$^{166}Er$ (b).

Figure 2: The Poincar\'e section of $^{160}Gd$ for
the energy $E=4$ MeV.

Figure 3: Poincar\'e sections
for $^{160}$Gd with $E=8$ MeV (below) and $E=18$ MeV (above).

Figure 4: Spectral statistics for $^{160}$Gd with
$6\leq E\leq 11$ MeV (below) and $15\leq E\leq 20$ MeV (above);
the solid lines are the Poisson statistics and the dashed lines are the
GOE statistics.

Figure 5: Poincar\'e sections for $^{166}$Er with
$E=4$ MeV (below) and $E=15$ MeV (above).

Figure 6: Spectral statistics for $^{160}$Er with
$2\leq E\leq 6$ MeV (below) and $13\leq E\leq 16$ MeV (above);
the solid lines are the Poisson statistics and the dashed lines are the
GOE statistics.

\newpage

\begin{center}
\begin{tabular}{|ccccc|} \hline\hline
Nucleus & $C_2$ (MeV) & $C_3$ (MeV) & $C_4$ (MeV) &
$V_0$ (MeV)\\ \hline
$^{160}Gd$ & -100.64 &  37.57 &  668.01 & 5.56\\
$^{166}Er$ &  -73.77 &  64.96 & 1138.52 & 1.89\\ \hline\hline
\end{tabular}
\end{center}
\vskip 0.6 truecm
{\bf Table 1}
\vskip 0.5 truecm

\end{document}